\documentclass[pra,aps,twocolumn,showpacs]{revtex4}

\usepackage{graphicx}

\begin{document}

\title{Optimal two-copy discrimination of quantum measurements}

\author{Jarom\'{\i}r Fiur\'{a}\v{s}ek} 
\affiliation{Department of Optics, Palack\'{y} University, 17. listopadu 12,
77900 Olomouc, Czech Republic}

\author{Michal Mi\v{cu}da} 
\affiliation{Department of Optics, Palack\'{y} University, 17. listopadu 12,
77900 Olomouc, Czech Republic}

\begin{abstract}
We investigate optimal discrimination between two  projective quantum measurements on a single qubit. We consider scenario where the measurement that should be identified can be performed twice and we show that adaptive discrimination strategy, entangled probe states, and feed-forward all help to increase the probability of correct identification of the measurement. We also experimentally  demonstrate the studied discrimination strategies and test their performance. The employed experimental  setup involves projective measurements on polarization states of single photons and preparation of required probe two-photon polarization states by the process of spontaneous parametric down-conversion and passive linear optics.

\end{abstract}


\maketitle

\section{Introduction}

Two non-orthogonal quantum states cannot be perfectly distinguished.
This fundamental constraint has important practical  consequences as it for instance guarantees the security 
of certain quantum key distribution protocols. Even if perfect discrimination is ruled out, one can nevertheless 
try to perform this task in an approximate manner. Various strategies for optimal approximate discrimination of quantum states
have been studied since the seminal work of  Holevo \cite{Holevo73} and Helstrom \cite{Helstrom76}. 
The rapid development of quantum information theory during recent years stimulated investigation of discrimination 
of more complex quantum objects, namely quantum operations, channels, and measurements \cite{Childs00,Acin01,DAriano05,Sacchi05,Wang06,Ji06,Ziman08,Ziman09,Li08,Chiribella08a,Duan09}. 
The role of  entanglement in discrimination of quantum operations and channels has been studied in some detail \cite{Dariano01,DAriano02,Duan07,Zhou07,Duan08,Piani09} and 
experimental realizations of several discrimination schemes for quantum operations have been reported \cite{Zhang08,Laing09}.
Very recently, quantum combs have been established as a general framework for treating the problems 
of discrimination and cloning of quantum operations \cite{Chiribella08,Chiribella09}.

Although formally the discrimination of quantum states and operations may look quite similar at first glance, there are  important differences due to the richer inherent  structure of quantum operations.
For instance, any two different unitary operations $U$ and $V$  
can be perfectly and deterministically discriminated provided that a sufficient finite number of applications of 
the operation is accessible \cite{Acin01}. Similarly, perfect discrimination between two different  projective quantum measurements is possible with finite number of uses of the measuring apparatus \cite{Ji06}. In contrast, two non-orthogonal quantum states cannot be perfectly deterministically discriminated from an arbitrary finite number of copies. 

In the present paper, we shall investigate in detail various strategies 
for discrimination among two quantum measurements \cite{Ji06,Ziman08,Ziman09}.
We shall assume that we are given a measuring apparatus $M$ that performs one of two single-qubit projective measurements $A$ or $B$. Our goal is to determine as well as possible whether $M=A$ or $M=B$ for a given fixed finite number of allowed utilizations of the measuring apparatus $M$.
In particular, we shall focus on the scenario where the measurement $M$ can be performed twice.
We will refer to this scenario as two-copy discrimination of quantum measurements. We will assume that no further auxiliary measurements could be performed on some ancilla states,
so the identity of the measurement has to be determined solely from the two outcomes of $M$. 
Already within this setting there exist several different discrimination strategies of varying complexity and performance.
We will show that adaptive discrimination, entangled probe states, and feed-forward all help to enhance the probability of correct identification of the measurement. 
By combining entangled probes and feed-forward, perfect deterministic 
 discrimination of projective measurements is possible provided that their distance is sufficiently large \cite{Ji06}. Here we explicitly derive the  entangled probe state and  feed-forward operation that enable perfect two-copy discrimination for a large class of pairs of single-qubit measurements. We also report on results of a successful 
proof-of-principle experimental realization of the studied discrimination strategies. 
We employ an optical setup where the goal is to  distinguish between two different projective measurements on a polarization state of single photon.

The rest of the present paper is organized as follows. In Sec. II we fix the notation and 
describe the various possible discrimination strategies. Sec. III is devoted to the analysis of discrimination
schemes employing single-qubit probes. Discrimination using entangled two-qubit probe states is treated in Sec. IV.
In Sec. V we describe the experimental implementation of the studied discrimination strategies 
and  discuss the experimental results. Finally, Sec. VI contains the conclusions.

\begin{figure}[!t!]
\centerline{\includegraphics[width=0.8\linewidth]{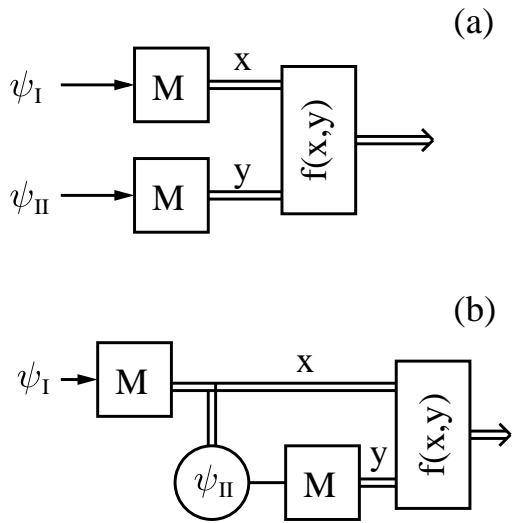}}
\caption{Discrimination of single-qubit quantum measurements with single-qubit probe states. In the considered scenario 
the measurement $M$ can be performed twice. 
(a) Discrimination using two fixed probe states $|\psi_{\mathrm{\mathrm{I}}}\rangle$ and $|\psi_{\mathrm{II}}\rangle$. (b) Adaptive discrimination strategy where the second probe
state $|\psi_{\mathrm{II}}\rangle$ is chosen according to the result of the first measurement $x$. Time flows from the left to the right. 
Single lines indicate quantum bits, double lines classical bits. }
\end{figure}

\section{Discrimination strategies}

Throughout the paper, the two single-qubit projective measurements that should be  discriminated will be labeled by letters $A$ and $B$. 
The two possible measurement outcomes will be denoted as $0$ and $1$, respectively.
Without loss of any generality we can choose the two projective measurements in the following form,
\begin{eqnarray}
\Pi_{A,0}&=&|0\rangle\langle 0|, \qquad \Pi_{B,0}=|\theta \rangle\langle \theta |, \nonumber \\
\Pi_{A,1}&=&|1\rangle \langle 1|, \qquad \Pi_{B,1}=|\theta_\perp \rangle \langle \theta_\perp|, 
\label{projectorsAB}
\end{eqnarray}
where 
\begin{equation}
|\theta\rangle = \cos\theta|0\rangle+\sin\theta|1\rangle, \quad
|\theta_\perp\rangle = \sin\theta|0\rangle-\cos\theta|1\rangle.
\end{equation}
 The angle $\theta\in[0,\pi/2]$ parameterizes the overlap $\mathcal{O}$ between the two measurements that can be naturally defined as
 \begin{equation}
 \mathcal{O}=\mathrm{Tr}[\Pi_{A,0}\Pi_{B,0}]=\mathrm{Tr}[\Pi_{A,1}\Pi_{B,1}]=\cos^2\theta.
 \end{equation}

We shall assume that the \emph{a-priori} probability of each measurement is $\frac{1}{2}$ and that no other auxiliary measurements could be performed. Although interesting phenomena arise mainly when several uses of the measurement are allowed, let us for the sake of completeness first consider the situation when the measurement can be performed only once.  A single-qubit probe state $|\psi\rangle$ is sent to the measuring apparatus and if the measurement outcome reads $0$ ($1$) then we guess that the measurement $A$ ($B$) was performed. The probability of correct guess $P_{\mathrm{succ}}$ is maximized if $|\psi\rangle$ is chosen as the eigenstate corresponding to the largest eigenvalue of operator $\Pi_{A,0}+\Pi_{B,1}$. We obtain 
\begin{equation}
|\psi\rangle = \frac{1}{\sqrt{2(1-\sin\theta)}} \left(\cos\theta|0\rangle+(\sin\theta-1)|1\rangle\right),
\end{equation}
and
\begin{equation}
P_{\mathrm{succ}}=\frac{1}{2}(1+\sin\theta).
\end{equation}

\begin{figure}[!t!]
\centerline{\includegraphics[width=0.8\linewidth]{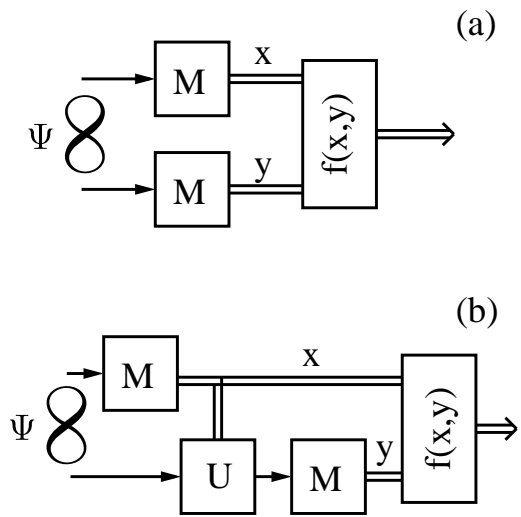}}
\caption{ (a) Discrimination of quantum measurements using entangled probe state $|\Psi\rangle$. 
(b) Feed-forward enhanced discrimination  of quantum measurements. A unitary operation $U$ depending 
on the measurement outcome $x$ on the first part of the entangled state is applied to the second part 
of the entangled state prior to the measurement.}
\end{figure}

Let us now assume that the measurement can be performed twice. In this case  we can distinguish four different discrimination strategies, as illustrated in Figs. 1 and 2.  The most straightforward approach is the probing by two fixed single-qubit states, see Fig. 1(a). The measurement is then inferred 
from the two measurement results  $x,y\in\{0,1\}$. The discrimination strategy is formally
described by a function $f(x,y)$ that assigns an estimate A or B to each of the four possible pairs of outcomes 00, 01, 10, 11.
There are altogether $2^4=16$ such functions and when determining the optimal discrimination strategy, we must optimize over all those 16 alternatives.  
  
The strategy shown in Fig. 1(a) can be improved by using an adaptive scheme, where the second single-qubit probe state
$|\psi_{\mathrm{II}}\rangle$ becomes dependent on the outcome $x$ of the measurement on the first probe state
$|\psi_\mathrm{I}\rangle$, cf. Fig. 1(b). There are thus two different second probe states $|\psi_{\mathrm{II},0}\rangle$ 
and $|\psi_{\mathrm{II},1}\rangle$ that 
can be optimized independently. As we shall show below, this adaptive procedure increases 
the probability of successful guess of the correct measurement.

So far we have considered probing by single-qubit states. A more general strategy, however, could explore an entangled two-qubit state
as a probe, as illustrated in Fig. 2(a). Moreover, we can combine the entanglement with feed-forward and after performing the measurement
on the first qubit of the entangled state we can apply to the second qubit a unitary operation $U(x)$ that depends on the
outcome of the first measurement \cite{Ji06}. This most advanced discrimination strategy is depicted in Fig. 2(b). This latter approach allows for perfect deterministic two-copy discrimination 
provided that $ \theta \geq \frac{\pi}{4}$.

\section{Probing with single-qubit states}
 
Let us first concentrate on the probing with two fixed single-qubit states as shown in Fig. 1(a). We choose as a figure  of merit  that should be maximized the probability of successful guess of the measurement,
\begin{equation}
P_{\mathrm{succ}}=\frac{1}{2}\sum_{x=0}^1 \sum_{y=0}^1 \mathrm{Tr}[\psi_{\mathrm{I}}\Pi_{f(x,y),x}] \mathrm{Tr}[\psi_{\mathrm{II}}\Pi_{f(x,y),y}].
\label{Psuccsingle}
\end{equation}
Here $\psi_j=|\psi_j\rangle \langle \psi_j|$ is a short-hand notation for a density matrix of a pure state. 
By convexity, the pure probe states are always optimal as can be directly seen from the structure 
of the formula (\ref{Psuccsingle}). For a fixed $\psi_{\mathrm{I}}$ and $f(x,y)$ the optimal $\psi_{\mathrm{II}}$ 
can be determined as the eigenstate of the operator
\begin{equation}
R=\frac{1}{2}\sum_{x=0}^1 \sum_{y=0}^1 \mathrm{Tr}[\psi_{\mathrm{I}}\Pi_{f(x,y),x}] \Pi_{f(x,y),y}
\label{Msingle}
\end{equation}
that corresponds to the maximum eigenvalue $r_{\mathrm{max}}$ of $R$. Maximizing $r_{\mathrm{max}}$ 
over all 16 functions $f(x,y)$ and over all probe states $\psi_{\mathrm{I}}$ then yields $P_{\mathrm{succ}}$. 
We have performed this optimization numerically and found that the optimal $f(x,y)$ is asymmetric, 
guess A is made for three outcomes and guess B is made only for one outcome,
\begin{equation}
f(0,0)=A, \quad f(0,1)=A, \quad f(1,0)=A, \quad f(1,1)=B. 
\label{fxysingle}
\end{equation}
Moreover, the optimal $\psi_{\mathrm{I}}$ and $\psi_{\mathrm{II}}$ lie in the same plane of the 
Bloch sphere as the projectors (\ref{projectorsAB}), which is intuitively  plausible. We can thus write 
\begin{eqnarray}
|\psi_{\mathrm{I}}\rangle&=&\cos \phi_I |0\rangle+\sin\phi_I|1\rangle, \nonumber \\
|\psi_{\mathrm{II}}\rangle&=&\cos \phi_{\mathrm{II}} |0\rangle+\sin\phi_{\mathrm{II}}|1\rangle.
\label{psiIandIIsingle}
\end{eqnarray}
Assuming the form (\ref{psiIandIIsingle}) of $\psi_{\mathrm{I}}$ and the guessing prescription (\ref{fxysingle}), 
the whole optimization can be performed analytically.
After some algebra one arrives at the expression for the optimal angle $\phi_I$,
\begin{equation}
\phi_{\mathrm{I}}=\frac{1}{2}\left(\theta-\arccos\left[\frac{1}{4\cos\theta}\left(1-\sqrt{1+8\cos^2\theta}\right)\right] \right).
\label{phiIfirststrategy}
\end{equation}
The corresponding success probability reads
\begin{eqnarray}
P_{\mathrm{succ,sep}}&=&\frac{1}{2}+\frac{\tan\theta}{8\sqrt{2}}\sqrt{1+2\cos(2\theta)+\sqrt{5+4\cos(2\theta)}}
\nonumber \\
&&+\frac{\sin\theta}{4\sqrt{2}}\sqrt{2+\cos(2\theta)+\sqrt{5+4\cos(2\theta)}}. \nonumber \\
\end{eqnarray}
Furthermore, it holds that it is optimal to set  $\phi_{\mathrm{II}}=\phi_{\mathrm{I}}$, hence the two optimal probe states $\psi_{\mathrm{I}}$ and $\psi_{\mathrm{II}}$ are in fact identical.

\begin{figure}[!t!]
\centerline{\includegraphics[width=0.98\linewidth]{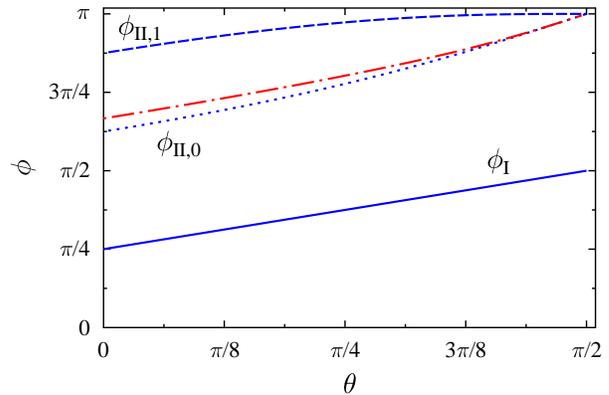}}
\caption{(Color online) The angles $\phi_j$ parameterizing optimal probe single-qubit states are plotted in dependence on $\theta$.
Shown are the optimal angles for adaptive strategy $\phi_{\mathrm{I}}$ (blue solid line),  $\phi_{\mathrm{II},0}$ (blue dotted line),
and $\phi_{\mathrm{II},1}$ (blue dashed line). Also shown is the optimal angle $\phi=\phi_{\mathrm{I}}=\phi_{\mathrm{II}}$ 
for strategy with fixed input probe states (red dot-dashed line).}
\end{figure}

Let us now move to the second scenario, where the second probe state $\psi_{\mathrm{II}}$ is chosen according to the result $x$ of the first measurement. The success probability of this protocol can be expressed as
\begin{equation}
P_{\mathrm{succ}}=\frac{1}{2}\sum_{x=0}^1\sum_{y=0}^1 \mathrm{Tr}[\psi_{\mathrm{I}}\Pi_{f(x,y),x}] \mathrm{Tr}[\psi_{\mathrm{II},x}\Pi_{f(x,y),y}].
\end{equation}
For each $f(x,y)$ and $\psi_{\mathrm{I}}$ the two probe states $\psi_{\mathrm{II},0}$ and $\psi_{\mathrm{II},1}$ have to be optimized independently.
They can be determined as eigenstates corresponding to maximum eigenvalues $r_{\mathrm{max},0}$ and $r_{\mathrm{max},1}$ of operators
\begin{eqnarray}
R_0&=&\sum_{y=0}^1 \mathrm{Tr}[\psi_{\mathrm{I}}\Pi_{f(0,y),0}] \Pi_{f(0,y),y}, \nonumber \\
R_1&=&\sum_{y=0}^1 \mathrm{Tr}[\psi_{\mathrm{I}}\Pi_{f(1,y),1}] \Pi_{f(1,y),y}.
\end{eqnarray}
The optimal discrimination strategy can be determined by maximizing $r_{\mathrm{max},0}+r_{\mathrm{max},1}$ over all $f(x,y)$
and $\psi_{\mathrm{I}}$. Numerical maximization reveals that, again, optimal $\psi$ all lie in the same plane as projectors (\ref{projectorsAB}) and 
are thus of the form (\ref{psiIandIIsingle}). The optimal function $f(x,y)$ is now symmetric,
\begin{equation}
f(0,0)=A, \quad f(0,1)=B, \quad f(1,0)=A, \quad f(1,1)=B.
\label{fpatternadaptive}
 \end{equation}
The optimization of the angles $\phi_{\mathrm{I}}$, $\phi_{\mathrm{II},0}$ and $\phi_{\mathrm{II},1}$ can be again performed fully analytically and we obtain
\begin{equation}
\phi_{\mathrm{I}}=\frac{\pi}{4}+\frac{\theta}{2}.
\label{phiIad}
\end{equation}
The  explicit formulas for the angles $\phi_{\mathrm{II},0}$ and $\phi_{\mathrm{II},1}$ are rather unwieldy and are not reproduced here.
Instead, we plot the dependence of the optimal $\phi_j$ on $\theta$ in Fig. 3. Note that generally $\phi_{\mathrm{II},0}\neq \phi_{\mathrm{II},1}$ which is a signature of the adaptive discrimination strategy.  The maximum achievable probability of success reads
\begin{equation}
P_{\mathrm{succ,ad}}=\frac{1}{2}\left(1+\sqrt{1-\cos^4\theta}\right).
\end{equation} 
It can be explicitly checked that for all $\theta \in (0,\pi/2)$ it holds that $P_{\mathrm{succ,ad}}>P_{\mathrm{succ,sep}}$
so the adaptive strategy strictly outperforms the strategy where the probe single-qubit states are a-priori fixed.

\section{Probing with entangled states}

We now switch our attention to protocols exploiting entangled two-qubit probe states. The first such scheme is shown in Fig. 2(a).
Here a measurement is performed on each qubit of a fixed two-qubit probe state $|\Psi\rangle$.
The success probability of this protocol can be written as
\begin{equation}
P_{\mathrm{succ}}=\frac{1}{2}\sum_{x=0}^1 \sum_{y=0}^1 \mathrm{Tr}[\Psi \Pi_{f(x,y),x}\otimes \Pi_{f(x,y),y}].
\end{equation}
This can be rewritten as $P_{\mathrm{succ}}=\mathrm{Tr}[\Psi R_{\mathrm{ent}}]$ where
\begin{equation}
R_{\mathrm{ent}}=\frac{1}{2}\sum_{x=0}^1 \sum_{y=0}^1  \Pi_{f(x,y),x}\otimes \Pi_{f(x,y),y}.
\end{equation}
The maximum achievable success probability can be thus determined by calculating the maximum eigenvalue $r_{\mathrm{ent,max}}$
of $R_{\mathrm{ent}}$ for all sixteen functions $f(x,y)$ and taking the maximum value. Since $R$ is a $4\times 4 $ matrix,
this optimization can be performed fully analytically.

It turns out that two different guessing strategies are optimal depending on the value of $\theta$. For $\theta\leq \theta_{\mathrm{th}}=\arccos \frac{1}{\sqrt{3}}$
the optimal function $f$ reads
\begin{equation}
f(0,0)=A, \quad f(0,1)=B, \quad f(1,0)=B, \quad f(1,1)=A.
\label{fABBA}
\end{equation}
The optimal (unnormalized) probe state has the form
\begin{equation}
|\Psi\rangle= \cos(2\theta)[|11\rangle-|00\rangle]+(1-\sin(2\theta))[|01\rangle+|10\rangle],
\label{PsientangledI}
\end{equation}
and the success probability  reads
\begin{equation}
P_{\mathrm{succ,ent}}=\frac{1}{2}[1+\sin(2\theta)].
\label{PsuccentangledI}
\end{equation}

For $\theta \geq \theta_{\mathrm{th}}$ the symmetry is broken and the optimal $f(x,y)$ is given by
\begin{equation}
f(0,0)=A, \quad f(0,1)=B, \quad f(1,0)=A, \quad f(1,1)=A.
\end{equation}
The corresponding optimal probe state can be expressed as
\begin{eqnarray}
|\Psi\rangle&=&|00\rangle-|11\rangle+\tan\theta |10\rangle  \nonumber \\
& & -\frac{\cos\theta \sqrt{3+\cos(2\theta)}}{\sqrt{2}+\sin\theta\sqrt{3+\cos(2\theta)}}|01\rangle ,
\end{eqnarray}
and yields a success probability 
\begin{equation}
P_{\mathrm{succ,ent}}=\frac{1}{2}\left(1+\sqrt{1-\cos^4\theta}\right),
\end{equation}
which coincides with $P_{\mathrm{succ}}$ achievable by adaptive strategy with single-qubit probes.

The most advanced among the studied strategies employs entanglement and feed-forward, as illustrated in Fig. 2(b). A measurement is performed on one part of the entangled two-qubit state and the measurement outcome $x$ determines the unitary operation performed on the second qubit before it is measured \cite{Ji06}. 
The two measurement results are then used to identify the measurement device as A or B according to a function $f(x,y)$.
Without any loss of generality we can assume that for $x=0$ the operation on the second qubit is an identity transformation while for $x=1$ a unitary operation $U$ is applied to the qubit. 
The success rate of the scheme can be expressed as
\begin{eqnarray}
P_{\mathrm{succ}}&=&\frac{1}{2}\sum_{y=0}^1 \mathrm{Tr}[\Psi \Pi_{f(0,y),0}\otimes \Pi_{f(0,y),y}] \nonumber \\
&&+\frac{1}{2}\sum_{y=0}^1 \mathrm{Tr}[\Psi \Pi_{f(1,y),1}\otimes U^\dagger\Pi_{f(1,y),y}U].
\end{eqnarray}
In order to determine the maximum achievable $P_{\mathrm{succ}}$ we have to calculate the maximum eigenvalue $r_{\mathrm{ff,max}}$ of the operator
\begin{equation}
R_{\mathrm{ff}}=\frac{1}{2}\sum_{y=0}^1 \left(\Pi_{f(0,y),0}\otimes \Pi_{f(0,y),y} +
\Pi_{f(1,y),1}\otimes U^\dagger\Pi_{f(1,y),y}U\right),
\end{equation}
and further maximize $r_{\mathrm{ff,max}}$ over all single-qubit unitary operations $U$. A thorough numerical optimization reveals that for $\theta<\frac{\pi}{4}$ the feed-forward does not provide any advantage and it is optimal to use the entangled state (\ref{PsientangledI}) without any feed-forward which leads to the success probability (\ref{PsuccentangledI}). 
The situation, however, changes dramatically for $\theta>\frac{\pi}{4}$. If $\theta \in[\frac{\pi}{4},\frac{\pi}{2}]$ then the two measurements can be perfectly and deterministically discriminated from two utilizations. The optimal $f(x,y)$ is given by Eq. (\ref{fABBA}). An analytical expression for the required  entangled probe state can be derived,
\begin{equation}
|\Psi\rangle=\alpha|00\rangle+\beta|10\rangle+\gamma |11\rangle,
\label{psientangledfeedback}
\end{equation}
where
\begin{eqnarray}
\alpha&=&\frac{1}{\sqrt{2}}\sqrt{1-\sqrt{1-\frac{1}{\tan^2\theta}}}, \nonumber \\
\gamma&=&-\frac{1}{\sqrt{2}\tan\theta}\sqrt{1+\sqrt{1-\frac{1}{\tan^2\theta}}}, \nonumber \\
\beta&=& -\frac{\alpha}{\tan\theta}-\gamma\tan\theta.
\end{eqnarray}
The conditional unitary operation $U$ on the second qubit that should be applied if the outcome of the first measurement reads $1$ is defined as follows,
\begin{eqnarray}
U |0\rangle=\frac{1}{\sqrt{\beta^2+\gamma^2}}(\gamma|0\rangle+\beta|1\rangle), \nonumber \\
U |1\rangle=\frac{1}{\sqrt{\beta^2+\gamma^2}}(-\beta|0\rangle+\gamma|1\rangle). 
\label{Ufeedforward}
\end{eqnarray}

\begin{figure}[t]
\centerline{\includegraphics[width=0.95\linewidth]{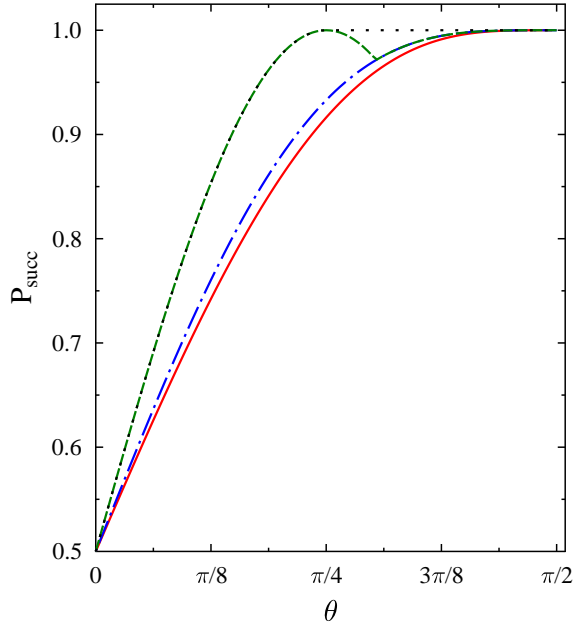}}
\caption{(Color online) Dependence of the success probability of the measurement discrimination $P_{\mathrm{succ}}$ on the angle $\theta$ is plotted for four different discrimination strategies: probing with fixed single qubit states (solid red line), adaptive strategy employing single-qubit states (dot-dashed blue line), probing with entangled state (dashed green line), and  combination of entangled probe state and a feed-forward (dotted black line).}
\end{figure}

The obtained results are illustrated in Fig. 4 that shows the dependence of $P_{\mathrm{succ}}$ on $\theta$ for the four studied discrimination strategies. We can see that adaptive strategy, entanglement and feed-forward all help to increase the success probability of the discrimination. For $\theta<\theta_{\mathrm{th}}$
the strategy based on a fixed entangled probe outperforms the adaptive strategy with single-qubit probe states. Interestingly, in the interval $\theta \in [\frac{\pi}{4},\theta_{\mathrm{th}}]$ the success probability 
of strategy involving fixed entangled probe decreases with increasing $\theta$, c.f. dashed green line in Fig.~4. This somewhat surprising feature arises because the class of discrimination strategies 
with fixed entangled probes represents
 only a subset of all possible strategies. By restricting ourselves to this class of strategies we impose
certain constraints which in this particular case give rise to the non-monotonicity of $P_{\mathrm{succ,ent}}$. If we employ a more general discrimination strategy combining entangled probe and
feed-forward then we recover  the intuitively expected monotonic dependence
of $P_{\mathrm{succ}}$ on $\theta$. Moreover, with this latter method
the two quantum measurements can be perfectly deterministically 
discriminated \cite{Ji06} when $\theta\geq \frac{\pi}{4}$.

\begin{figure}[b]
\centerline{\includegraphics[width=0.95\linewidth]{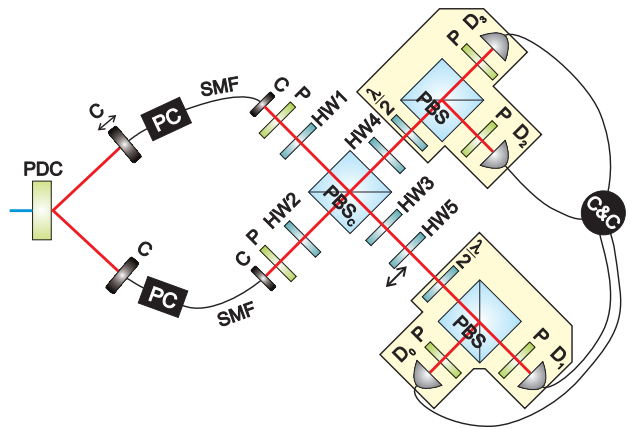}}
\caption{(Color online) Experimental setup. The scheme  consists of a nonlinear crystal where pairs of
photons are generated in the process of spontaneous parametric down conversion (PDC),
single-mode fibers (SMF), fiber polarization controllers (PC), fiber in/out couplers (C), bulk polarizers (P), 
half-wave plates (HW, $\lambda/2$), polarizing beam splitters (PBS), single-photon detectors (D), and coincidence electronics (C\&C).
The half-wave plate HW5 was inserted in the setup only for measurements with entangled probe states.}
\end{figure}

\section{Experiment}

In order to test the above developed discrimination strategies, we have experimentally implemented discrimination of
 projective measurements on a polarization state of a single photon. 
In the experiment, the alternative A corresponds to the measurement in the basis of horizontally/vertically polarized single-photon states, ($|H\rangle$, $|V\rangle$), while the alternative B represents a measurement in the basis of linearly polarized states rotated by angle $\theta$ with respect to the $H/V$ basis, ($\cos\theta|H\rangle+\sin\theta|V\rangle$, $\sin\theta|H\rangle-\cos\theta|V\rangle$).
As shown in Fig. 5, the measurement block consists of a half-wave plate ($\frac{\lambda}{2}$) 
whose rotation angle defines the measurement basis, polarizing beam splitter (PBS), two polarizers (P)
and two avalanche photodiodes serving as single-photon detectors (D). The  polarizers in front of the detectors filter out any possible remaining undesired signal that could be present due to imperfections of the PBS. Similarly as in Ref. \cite{Laing09}, we make use of spatial multiplexing, where two physical copies of the 
measuring apparatus are available. Note that this is merely a technical simplification of the experiment and all the developed discrimination strategies can be implemented also with only a single apparatus by using time multiplexing and delay lines.

In our experiment, pairs of temporally correlated horizontally polarized photons are  generated in the process of frequency degenerate Type-I non-collinear parametric downconversion in a LiIO$_3$ crystal pumped by a cw laser diode emitting $40$ mW of power  at the wavelength of $407$ nm \cite{Micuda08}. The downconverted signal and idler photons at wavelength $814$ nm are spatially filtered by coupling them
into single mode fibers.  After release into free space, the polarization states of the photons can be set and controlled by polarizers P and half-wave plates HW1 and HW2. The photons then impinge onto a polarizing beam splitter PBS$_C$ and propagate through additional half-wave plates HW3, HW4 (and, optionally, HW5) before impinging onto the two detection blocks. 
The central polarizing beam splitter PBS$_C$ together with the wave plates provide sufficient flexibility necessary for the implementation of the various discrimination strategies including those which require preparation of specific entangled states.

We begin by implementation of the strategy employing two equally polarized single-photon probes.
The wave plates HW1 and HW2 are set to $0^{\circ}$ such that both signal and idler photons are horizontally polarized and are fully transmitted through PBS$_{C}$. The half-wave plates HW3 and HW4 are rotated such as to prepare both photons in pure linear polarization state $|\psi\rangle=\cos\phi_{\mathrm{I}}|H\rangle+\sin\phi_{\mathrm{I}}|V\rangle$,  where 
the angle $\phi_{\mathrm{I}}$ is given by Eq. (\ref{phiIfirststrategy}). The photons are then detected by the measurement blocks and the coincidences between clicks of one detector from each block are recorded. 
Following Eq. (\ref{fxysingle}), if the coincidences D0\&D2, D0\&D3, or D1\&D2 are observed then we guess that the apparatus performs measurement in the H/V basis (device A), while if the coincidence D1\&D3 is recorded then we conclude that the apparatus performs measurement in the rotated basis (device B). We measure the coincidences for both basis settings and from the experimental data we calculate the success probability of a correct guess assuming that the \emph{a-priori} probability of each measurement device A or B was $\frac{1}{2}$. 

We then proceed to the adaptive discrimination strategy. In the present proof-of-principle experiment, we were not able to realize real-time adaptive measurement strategy, but we nevertheless successfully emulated this approach as follows. For both devices A and B (i.e. both basis settings) we perform two measurements. First, the signal photon is prepared by HW3 in a linearly polarized state at angle $\phi_{\mathrm{I}}=\frac{\pi}{4}+\frac{\theta}{2}$, cf. Eq. (\ref{phiIad}), the idler photon is prepared by HW4 in a linearly polarized state $\cos\phi_{\mathrm{II},0}|H\rangle+\sin\phi_{\mathrm{II},0}|V\rangle$, and the four coincidences are measured. The second measurement is almost identical to the first one except that  the idler photon is prepared in a state $\cos\phi_{\mathrm{II},1}|H\rangle+\sin\phi_{\mathrm{II},1}|V\rangle$. From the first (second) measurement we take into account only coincidences D0\&D2 and D0\&D3 (D1\&D2 and D1\&D3). The success probabilities are then calculated from this combined experimental data and according to the identification pattern (\ref{fpatternadaptive}).

The entanglement-based strategies are much more experimentally demanding because the two photons have to be prepared in an entangled state whose quality depends on the visibility of two-photon interference on PBS$_C$. First we address the simpler strategy without feed-forward. The probe state (\ref{PsientangledI}) is actually maximally entangled and can be rewritten as,
\begin{eqnarray}
|\Psi\rangle&=&|H\rangle[(\cos\theta-\sin\theta)|V\rangle -(\cos\theta+\sin\theta)|H\rangle] \nonumber \\
& &+|V\rangle[(\cos\theta+\sin\theta)|V\rangle+(\cos\theta-\sin\theta)|H\rangle]. \nonumber \\
\label{psientangledoptical}
\end{eqnarray}
 We rotate HW1 and HW2 by $22.5^{\circ}$ to prepare both signal and idler photons in front of PBS$_C$ in diagonally polarized state $\frac{1}{\sqrt{2}}(|H\rangle+|V\rangle)$. The two-photon state right at the output of PBS$_C$ conditional on a single photon propagating in each arm is maximally entangled and reads $\frac{1}{\sqrt{2}}(|HH\rangle+|VV\rangle)$.
This state can be transformed into the desired state (\ref{psientangledoptical}) by rotating the wave plates HW3 and HW4. In order to compensate for an unwanted $\pi$ phase shift we also insert an additional half-wave plate HW5 into the setup. We  measure coincidences for both measurement bases and determine the probability of successful discrimination from the acquired data. 

Finally, we  test the strategy involving entangled states and feed-forward. Since a fast feed-forward loop was not at our disposal we have decided to emulate this strategy similarly as in the case of adaptive strategy. We have determined setting of the wave plates HW1--HW5
which yields the partially entangled two-photon probe state (\ref{psientangledfeedback}). For a given fixed $\theta$ we measure coincidences for both measurement bases and then we  
rotate half-wave plate HW4 such that this operation is equivalent to the feed-forward transformation $U$, cf. Eq. (\ref{Ufeedforward}). We repeat all measurements for this altered configuration. 
From the first set of data we extract coincidences D0\&D2 and D0\&D3 and from the second set of data we  use coincidences D1\&D2 and D1\&D3. This yields the same data as a true feed-forward scheme where the rotation of the wave plate HW4 is performed only when detector D1 clicks and before the idler photon passes through HW4.

\begin{figure}[!t!]
\centerline{\includegraphics[width=0.95\linewidth]{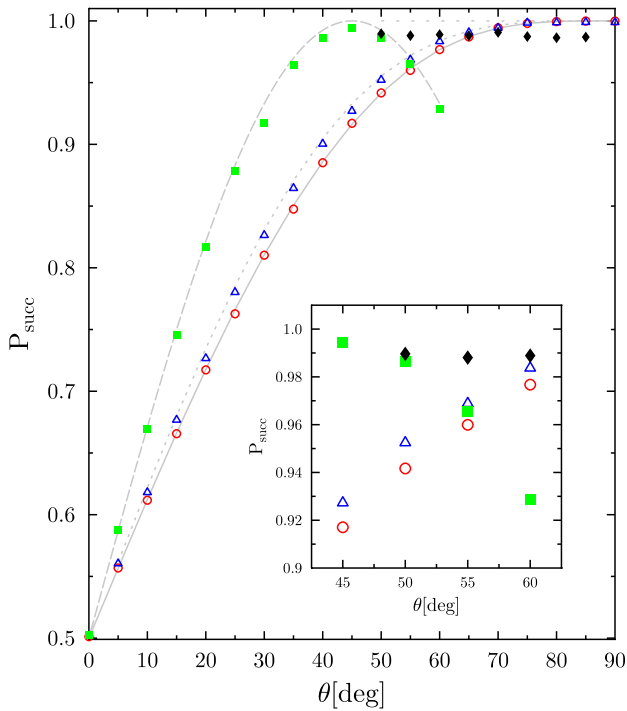}}
\caption{(Color online) Experimental results. Experimentally determined probability of successful discrimination $P_{\mathrm{succ}}$ is plotted for four different discrimination strategies: strategy with fixed single-qubit probes  (red empty circles), adaptive strategy with single qubit probes (empty blue triangles), 
strategy employing  two-qubit entangled state (\ref{psientangledoptical}) (green squares, measurements made for $0^{\circ}\leq\theta\leq 60^\circ$)  and strategy combining entanglement and feed-forward 
(black diamonds, measurements made for $50^{\circ}\leq\theta\leq 85^\circ$). The thin grey lines show the corresponding theoretical curves. The inset shows details around $\theta=55^\circ$.}
\end{figure}

The experimentally determined success probabilities for all four strategies are shown in Fig. 6. The results agree very well with the theoretical predictions. The statistical error of the measured $P_{\mathrm{succ}}$ is below $2\times 10^{-3}$. The error bars are smaller than the sizes of the symbols used in the graph and are thus not plotted. As predicted, the adaptive scheme outperforms the scheme with fixed single-qubit probe states, and the use of entangled states further significantly improves the probability of successful discrimination. We can see that for $\theta>\theta_{\mathrm{th}}\approx 54.7^{\circ}$
the entangled state (\ref{PsientangledI}) ceases to be optimal as expected. The discrepancy between theory and experiment is larger for entanglement-based strategies than for strategies with separable probes, because the performance of the former is affected by less-than unit visibility of two-photon interference on PBS$_C$ (we measure $V=0.98$) and the imperfections of PBS$_C$. Nevertheless, the strategy combining entanglement and feed-forward consistently achieves success probability $\approx 99\%$ for
 $\theta\in(50^{\circ},85^{\circ})$. In particular, for $\theta=55^\circ$ the advantage of using 
the entanglement and feed-forward is clearly visible from the experimental results, cf. inset in Fig. 6. As $\theta$ approaches $90^{\circ}$ the strategies involving separable probe states eventually 
outperform the entanglement-based strategy because they are much less affected by the technical imperfections.

\section{Conclusions}

In the present paper we have studied the discrimination between two projective single-qubit quantum measurements. We have seen that if two applications of the measurement are possible, then there exist several different discrimination strategies of varying complexity and performance. We have found that adaptive strategy, entanglement and feed-forward all help to increase the success probability of the discrimination. 
We have explicitly determined an entangled probe state that, together with feed-forward,  
enables perfect deterministic two-copy discrimination of the two  measurements for $\frac{\pi}{4}\leq \theta\leq \frac{\pi}{2}$. This is analogous to the perfect finite-copy distinguishability  of unitary transformations and general quantum operations \cite{Acin01,Ji06,Duan09}.
We have successfully experimentally confirmed the performance of the investigated protocols using a linear optical scheme where the task was to discriminate between two different measurements on a polarization state of single photon.

The discrimination scenarios considered in this paper did not involve any other ancilla measurements in addition to the measurement $M$ that should be discriminated. If we allow for such additional ancilla measurements then we can in principle construct even more sophisticated discrimination schemes that can be described by the formalism of quantum combs \cite{Chiribella08,Chiribella09}. Investigation of such advanced strategies as well as strategies involving inconclusive outcomes \cite{Ziman08} is left for future work. On the experimental side, we are currently working on a fast feed-forward loop that would allow us to fully implement the adaptive and feed-forward-based schemes.

\begin{acknowledgments}
This work was supported by Research Projects  ``Center of Modern
Optics'' (LC06007) and ``Measurement and Information in Optics''
(MSM6198959213) of the Czech Ministry of Education and by GACR (202/09/0747).

\end{acknowledgments}

\end{document}